\documentclass{PoS}
\usepackage{graphicx}
\usepackage{dcolumn}
\usepackage{bm}

\newcommand{\ket}[1]{\ensuremath{| {#1} \rangle }}
\newcommand{\bra}[1]{\ensuremath{\langle {#1} |}}
\newcommand{\be}{\begin{equation}}
\newcommand{\ee}{\end{equation}}
\newcommand{\beq}{\begin{equation}}
\newcommand{\eeq}{\end{equation}}
\newcommand{\bea}{\begin{eqnarray}}
\newcommand{\eea}{\end{eqnarray}}

\newcommand{\beqn}{\begin{eqnarray}}   
\newcommand{\eeqn}{\end{eqnarray}}

\renewcommand{\vec}[1]{\bm{#1}}

\title{Real photon emissions in leptonic decays}

\ShortTitle{Real photon emissions in leptonic decays}

\author{\speaker{G. Martinelli}\thanks{guido.martinelli@roma1.infn.it} , M.  Di Carlo, F. Mazzetti
        \\ Physics Department and INFN Sezione di Roma La Sapienza, Piazzale Aldo Moro 5, 00185 Roma, Italy
        }

\author{G.M.~de~Divitiis, A.~Desiderio, R.~Frezzotti, M.~Garofalo,
M.~Hansen, N.~Tantalo
\\ University of Rome Tor Vergata and INFN Roma Tor Vergata,
Via della Ricerca Scientifica 1, I-00133, Rome, Italy }
\author{D.Giusti, V.~Lubicz \\ Dip. di Matematica e Fisica, University of   Roma Tre and INFN   Roma Tre,\\ Via della Vasca Navale 84, I-00146 Rome, Italy}
\author{C.T.~Sachrajda \\Department of Physics and Astronomy, University of Southampton,\\ Southampton SO17 1BJ, UK}
\author{ F.~Sanfilippo, S.~Simula\\  INFN   Roma Tre,\\ Via della Vasca Navale 84, I-00146 Rome, Italy}
\abstract{We present a non-perturbative calculation of the form factors which contribute to the amplitudes for the radiative decays $P\to \ell \bar \nu_\ell   \gamma$, where $P$ is a pseudoscalar meson and $\ell$ is a charged lepton.   Together with the non-perturbative determination of the virtual photon corrections  to the processes $P\to \ell \bar \nu_\ell$, this will allow  accurate predictions to be made at $O(\alpha_{em})$  for  leptonic decay rates for pseudoscalar mesons ranging from the pion to the $B$ meson.  We are able to separate unambiguously the point-like contribution, the square of which leads to the infrared divergence in the decay rate,  from the structure dependent, infrared-safe, terms in the amplitude.   The fully non-perturbative, $O(a)$ improved  calculation  of the inclusive leptonic decay rates will lead to significantly improved  precision in the determination of the corresponding Cabibbo-Kobayashi-Maskawa  (CKM) matrix elements. Precise predictions for the emission of a hard photon are also very interesting, especially for the decays of heavy $D$ and $B$ mesons for which currently only model-dependent predictions are available to compare with existing experimental data.}

\FullConference{37th International Symposium on Lattice Field Theory - Lattice2019\\
		16-22 June 2019\\
		Wuhan, China}

\begin{document}

\section{Introduction}
The determination of the  CKM matrix  elements  represents a crucial test of the limits of the Standard Model (SM) in the quest for new physics.  In this framework, precise  experimental measurements and accurate  theoretical  predictions of the leptonic decay rates of light and heavy pseudoscalar mesons  are particularly  important.  A first-principles calculation of these quantities requires the non-perturbative determination of the physical amplitudes/rates that can only be obtained from QCD simulations on the lattice. In order to fully exploit the available experimental information~\cite{Tanabashi:2018oca}, strong isospin-breaking effects and  $O(\alpha_{em})$ electromagnetic corrections must be included. In particular,  we must be able to compute the rates for  radiative leptonic decays $P   \to \ell \bar \nu_\ell \gamma$, where $P$ is a  charged pseudoscalar meson, $\gamma$ a photon, $\ell$ a (anti-)lepton and $\bar \nu_\ell$ the corresponding  anti-neutrino (neutrino). This would also allow accurate, model-independent  predictions of  the important radiative decays of heavy mesons  with the emission of a hard photon.   Results of a lattice calculation of real-photon emission amplitudes have also been presented at this conference in ref.~\cite{Kane:2019jtj}.

In the limit of soft-photon energies,  the radiative decay rate can be reliably calculated perturbatively by treating the meson  as a point-like particle.  This  limit is however an idealisation and experimental measurements  are inclusive up to photon energies that might be too large to neglect  structure-dependent (SD) corrections to the point-like approximation.    
The region of hard  photon energies,  which   is particularly important for heavy mesons,  represents a fundamental probe of the internal structure of the mesons and  can only be studied in  lattice QCD simulations.  On the other hand,  even in the case of light mesons, where  chiral perturbation theory can be used, the low-energy constants entering  at $O(p^6)$ can only  be estimated  using  model-dependent assumptions~\cite{Bijnens:1996wm}-\cite{Cirigliano:2011ny}.    

In ref.~\cite{Carrasco:2015xwa}  a strategy to compute  QED radiative corrections to the $P\to \ell \bar \nu_\ell (\gamma)$ decay rates at $O(\alpha_{em})$ on the lattice was proposed. The strategy was  subsequently  applied  to provide the first non-perturbative model-independent calculation of the decay rates $\pi^-\to \mu^- \bar \nu_\mu (\gamma)$ and  $K^-\to \mu^- \bar \nu_\mu (\gamma)$~\cite{Lubicz:2016xro}-\cite{DiCarlo:2019thl}. The real soft-photon contributions was calculated in the point-like effective theory and the SD corrections were estimated,   by relying on the quoted chiral perturbation theory results, to be negligible (see~\cite{Carrasco:2015xwa}). On the other hand  SD corrections might  be relevant for the decays of pions and kaons into electrons when the energy of the photon becomes larger than  about $20$~MeV. Moreover, in the  single-pole dominance approximation proposed  in ref.~\cite{Becirevic:2009aq},  the SD contribution  was estimated to be rather large  in the case of heavy flavours. This contribution can  be precisely determined only in lattice QCD.   
Here we present  a non-perturbative, $O(a)$ improved  lattice calculation of the form factors entering the radiative decay rate $P\to \ell \bar \nu_\ell \gamma$ in the case of pions, kaons, $D$ and $D_s$ mesons. The case of bottom mesons will be studied in a future work  on the subject.
\section{Form factors contributing to the radiative decay amplitude}
The non-perturbative  hadronic amplitude for  the  process $P\to \ell\nu_\ell \gamma$ is given by the T-product 
\beq
H^{\alpha r}_W(k,p)=\epsilon_\mu^r(k)\, H^{\alpha \mu}_W(k,p)
=
\epsilon_\mu^r(k)\, \int d^4 y\, e^{ik\cdot y}\, \bra{0} \mathtt{T}\{ j_W^\alpha(0) j^\mu_{em}(y)\}\ket{P(\vec p)}\;,
\label{eq:starting}
\eeq
where $\epsilon_\mu^r(k)$ is the polarization vector of the photon with four-momentum $k$, $j^\mu_{em}$ is the electromagnetic  current, $j_W^\alpha$ is the hadronic weak current, $j_W^\alpha=V^\alpha-A^\alpha = \bar q_1 \, (\gamma^\alpha - \gamma^\alpha \gamma_5) \, q_2$, and $\vec p$ is the momentum of the meson $P$ with  mass  $m_P$.  To this amplitude, at $O(\alpha_{em})$,  we have to add the diagram in which the photon is emitted from the final-state charged lepton.  The latter contribution can however, be computed  in perturbation theory using the meson decay constant $f_P$. 
The decomposition of $H^{\alpha r}_W(k,p)$ in terms of form-factors has been discussed, for example,  in refs.~\cite{Carrasco:2015xwa,Bijnens:1992en}
\bea
&& H^{\alpha r}_W(k,p) =
\epsilon_\mu^r(k)\Bigg\{
H_1\,\left[k^2 g^{\mu\alpha}-k^\mu k^\alpha\right]
+
H_2\, \left[(p\cdot k-k^2)k^\mu-k^2(p-k)^\mu\right](p-k)^\alpha  \label{eq:ffdef}
 \\
& &
-i\frac{F_V}{m_P}\varepsilon^{\mu\alpha\gamma\beta}k_\gamma p_\beta
+\frac{F_A}{m_P}\left[(p\cdot k-k^2)g^{\mu\alpha}-(p-k)^\mu k^\alpha\right]
+
f_P\left[g^{\mu\alpha}+\frac{(2p-k)^\mu(p-k)^\alpha}{2p\cdot k-k^2}\right]
\Bigg\}\;. \nonumber
\eea
The last term in  Eq.~(\ref{eq:ffdef}) corresponds to the point-like infrared-divergent contribution. This term saturates the Ward Identity satisfied by $H^{\alpha \mu}_W(k,p)$, i.e.
$
k_\mu\, H^{\alpha \mu}_W(k,p)= i\bra{0}j_W^\alpha(0)\ket{P(p)} = f_P\, p^\alpha\;.
$
  The four form-factors $H_{1,2}$ and $F_{V,A}$ are scalar functions of Lorentz invariants, the squared meson mass $m^2_P$,  $p\cdot k$ and $k^2$.   Eq.~(\ref{eq:ffdef}) is valid for generic (off-shell) values of the photon momentum and for generic choices of the polarisation vectors. By setting the photon on-shell, i.e. by taking $k^2 =0$,  at fixed meson mass   the form factors are functions of  $p\cdot k$ only.  A convenient dimensionless variable is given by $x_\gamma=2p\cdot k /m_P^2$.   By choosing a \emph{physical} basis for the polarization vectors such that  $\epsilon_r\cdot k=0$   we have
\beq
H^{\alpha r}_W(k,p) =
\epsilon_\mu^r(k)\Bigg\{
-i\frac{F_V(x_\gamma)}{m_P}\varepsilon^{\mu\alpha\gamma\beta}k_\gamma p_\beta
+
\left[\frac{F_A(x_\gamma)}{m_P}+\frac{f_P}{p\cdot k}\right]
\left(p\cdot k\, g^{\mu\alpha}-p^\mu k^\alpha\right)
+
\frac{f_P}{p\cdot k}\, p^\mu p^\alpha
\Bigg\}\;.
\label{eq:senzalabel}
\eeq
Once the decay constant $f_P$ and the two SD axial and vector form-factors $F_A$ and $F_V$ are known, the  decay rate can be calculated by using the formulae given in \cite{Bijnens:1992en} and in appendix B of ~\cite{Carrasco:2015xwa}.
\section{Extracting the form  factors from  Euclidean correlators}
The Euclidean correlation function corresponding to Eq.~(\ref{eq:starting}) is given by 
\beq
C^{\alpha r}_W(t,\vec p, \vec k)=- i \, \epsilon_\mu^r(\vec k)  \, \int d^4 y\,\int d^3 \vec x\, \, \langle  0\vert \mathtt{T}\{ j_W^\alpha(t,\vec 0) j^\mu_{em}(y)\}P(0,\vec x)\vert 0 \rangle \,e^{E_\gamma t_y-i\vec k\cdot \vec y+ i \vec p \cdot \vec x}\,  
\label{eq:correlator}
\eeq
where   $k=(iE_\gamma,\vec k)$, with $E_\gamma=\vert \vec k \vert$,  $p=(iE,\vec p)$ and $\int d^3 \vec x\, P(0,\vec x)\,e^{i \vec p \cdot \vec x}$ is the source of the pseudoscalar meson with momentum $\vec p$. The convergence of the integral over $t_y$ is ensured by the safe analytic continuation from Minkowski to Euclidean space, because of the absence of intermediate states lighter than the pseudoscalar meson. The physical form factors can be extracted directly from the Euclidean correlation functions
\beq
R^{\alpha r}_W(t;\vec p,\vec k) 
= \frac{2E}{e^{-t(E-E_\gamma)}\, \bra{P(\vec p)}P\ket{0}}\, C^{\alpha r}_W(t;\vec p,\vec k)
=
H^{\alpha r}_W(k,p) + \cdots
\label{eq:Rinf}
\eeq
where $ \bra{P(\vec p)}P\ket{0}$ is the matrix element of the operator $P$ between the vacuum and the meson state and the dots represent sub-leading exponentials. 
It is useful to note that, in order to separate the axial and vector form-factors it is enough to compute separately the ratios $R^{\alpha r}_{V,A}(t;\vec p,\vec k)$ corresponding to the (renormalised) vector and axial component of the weak current, see eq.~(\ref{eq:estimators}) below. For $j_{em}^\mu$ an exactly conserved lattice vector current is employed.
\begin{figure}[!t]
\begin{center}
\includegraphics[width=0.30\textwidth]{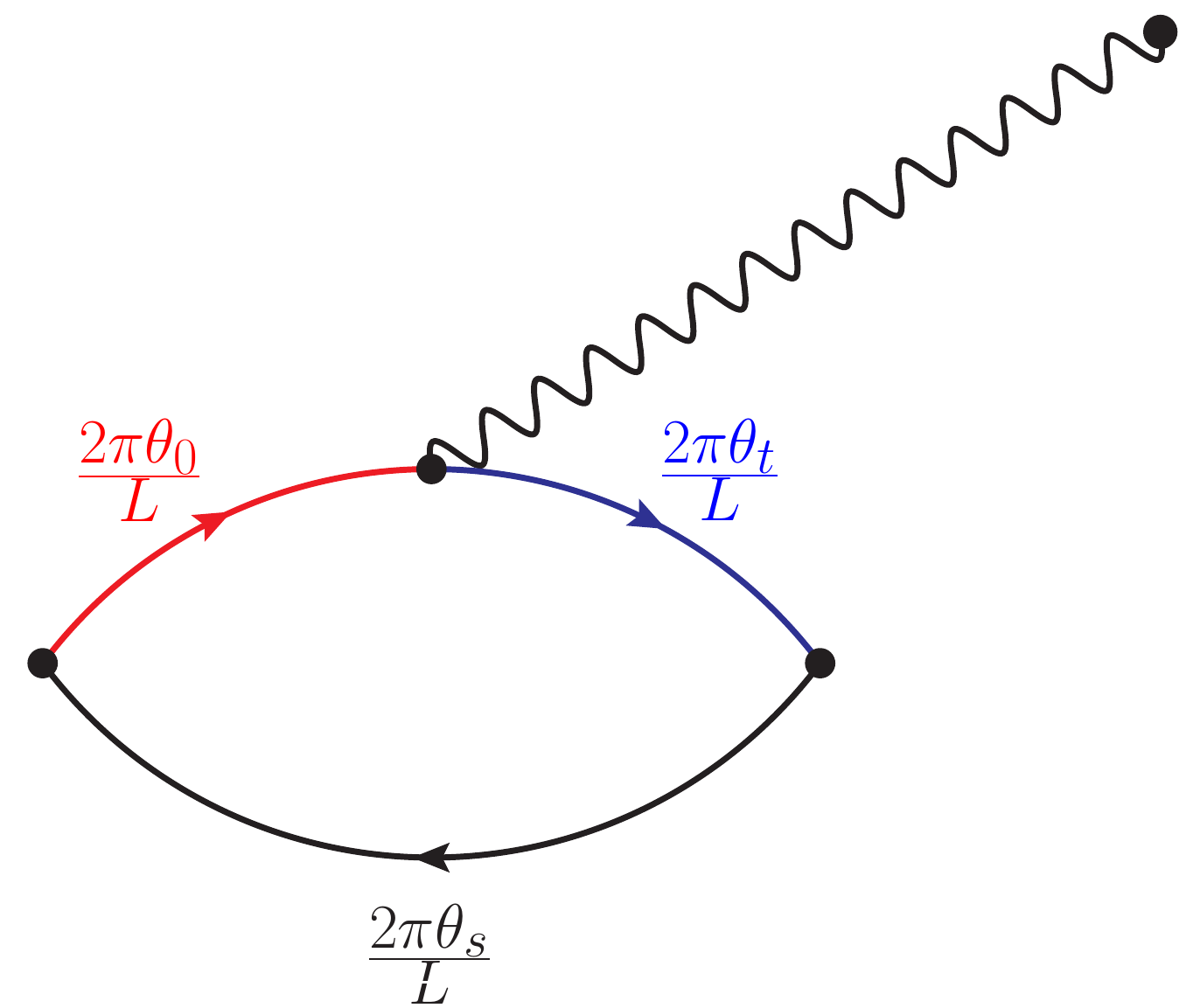}\,\,\,\,\,\,\,\,\,\,\,\,\,\,\,\,\,\,\,\,\,\,\,\,\,\,\,\,\,\,\,\,\,\,\,\,\,\,
\includegraphics[width=0.30\textwidth]{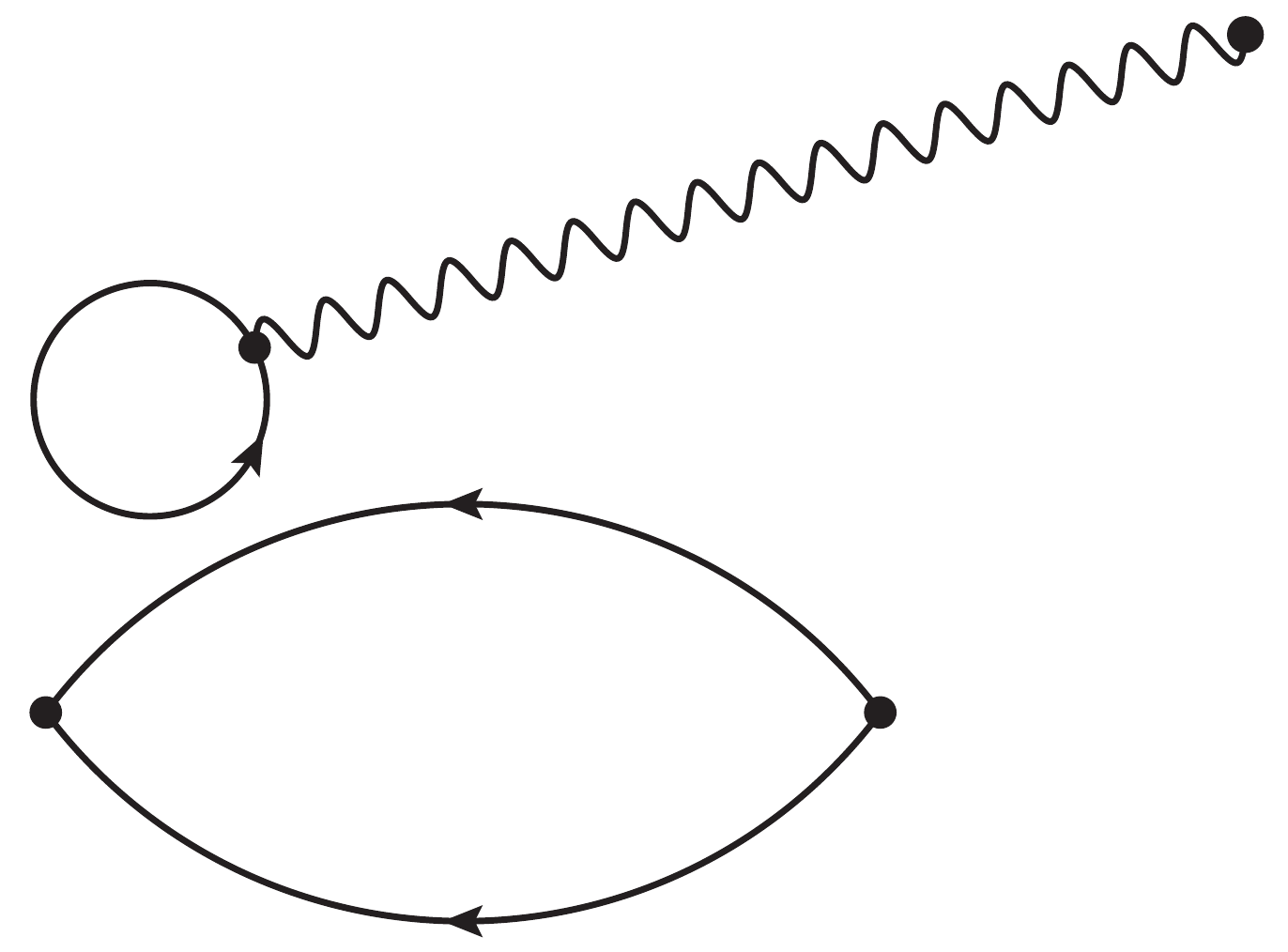}
\end{center}
\caption{
\footnotesize  The {\it connected} diagram on the left shows  our choice of the spatial boundary conditions. By treating the two propagators attached to the electromagnetic current  as two different flavours, with  the same mass and electric charge but different boundary conditions, we may  choose arbitrary values for the meson and photon spatial momenta.The diagram on the right represents the contribution associated with the emission of the photon by the sea-quarks. By neglecting  this diagram we have been working  in the so-called electro-quenched approximation.}
\label{fig:disctheta}
\end{figure}
The previous discussion assumed an infinite time extent ($T$) of the lattice. In our numerical calculations we have employed numerical estimators for the ratios $R^{\alpha r}_{V,A}(t;\vec p,\vec k)$ built in terms of finite--$T$ correlators that properly  account  for the fact that the simulated quark and gauge fields satisfy respectively anti-periodic and periodic boundary conditions in time. 

Within the electro-quenched approximation, i.e. in  the absence of the disconnected contribution  shown in the right-panel of Fig.~\ref{fig:disctheta}, it is possible to choose arbitrary values of the spatial momenta by using different spatial boundary conditions~\cite{deDivitiis:2004kq,Sachrajda:2004mi} for the quark fields. More precisely, we set the boundary conditions for the ``spectactor'' quark such that $\psi(x+\vec{\hat k} L)=\exp(2\pi i \vec{\hat k} \cdot \vec \theta_s/L) \psi(x)$. Then we treat the two propagators that are connected with the electromagnetic current (the red and blue lines) as the results of the Wick contractions of two different fields having the same mass and electric charge but satisfying different boundary conditions. This is possible at the price of accepting tiny violations of unitarity that are exponentially suppressed in  the volume (similar effects are induced in any case by the electro-quenched approximation). By setting the boundary conditions  as illustrated in the figure,  we have thus been able to choose arbitrary values for the meson and photon spatial momenta,  $\vec p = \frac{2\pi}{L}\left( \vec \theta_0-\vec \theta_s\right)$ and   $\vec k = \frac{2\pi}{L}\left( \vec \theta_0-\vec \theta_t\right)$   by tuning the real three-vectors $\vec \theta_{0,t,s}$, where the subscript $i=0,t,s$ in the definition  refers  to    the quark line emerging from  the source  in the origin, $0$; the quark   annihilating in the sink given by the hadronic weak current at  time $t$ and    the {\it spectator quark} respectively.   The numerical results  have been obtained by choosing all the non-zero components of the spatial momenta to be  along the z-direction, i.e.  $\vec p=(0,0,\vert \vec p\vert)$ and $\vec k=(0,0,E_\gamma)$.   With this particular choice a convenient basis for the polarization vectors of the photon  is the one in which the \emph{two} physical polarization vectors are given by
$  \epsilon^\mu_{1,2}=\left(0, \mp \frac{1}{\sqrt{2}},  -\frac{1}{\sqrt{2}}, 0\right)$. In this basis we have
$\epsilon_r\cdot p =\epsilon_r\cdot k=0$  
and  consequently
\beq
H^{jr}_A(k,p)=
\frac{\epsilon^j_r\, m_P}{2}\, x_\gamma\left[F_A(x_\gamma)+\frac{2f_P}{m_P x_\gamma}\right]\;,
\qquad
H^{jr}_V(k,p)=\frac{i \left(E_\gamma\, \vec \epsilon_r \wedge \vec p-E\, \vec \epsilon_r \wedge \vec k\right)^j}{m_P}\, 
F_V(x_\gamma)\;.
\eeq
\begin{figure}[!t]
\begin{center}
\includegraphics[width=0.45\textwidth]{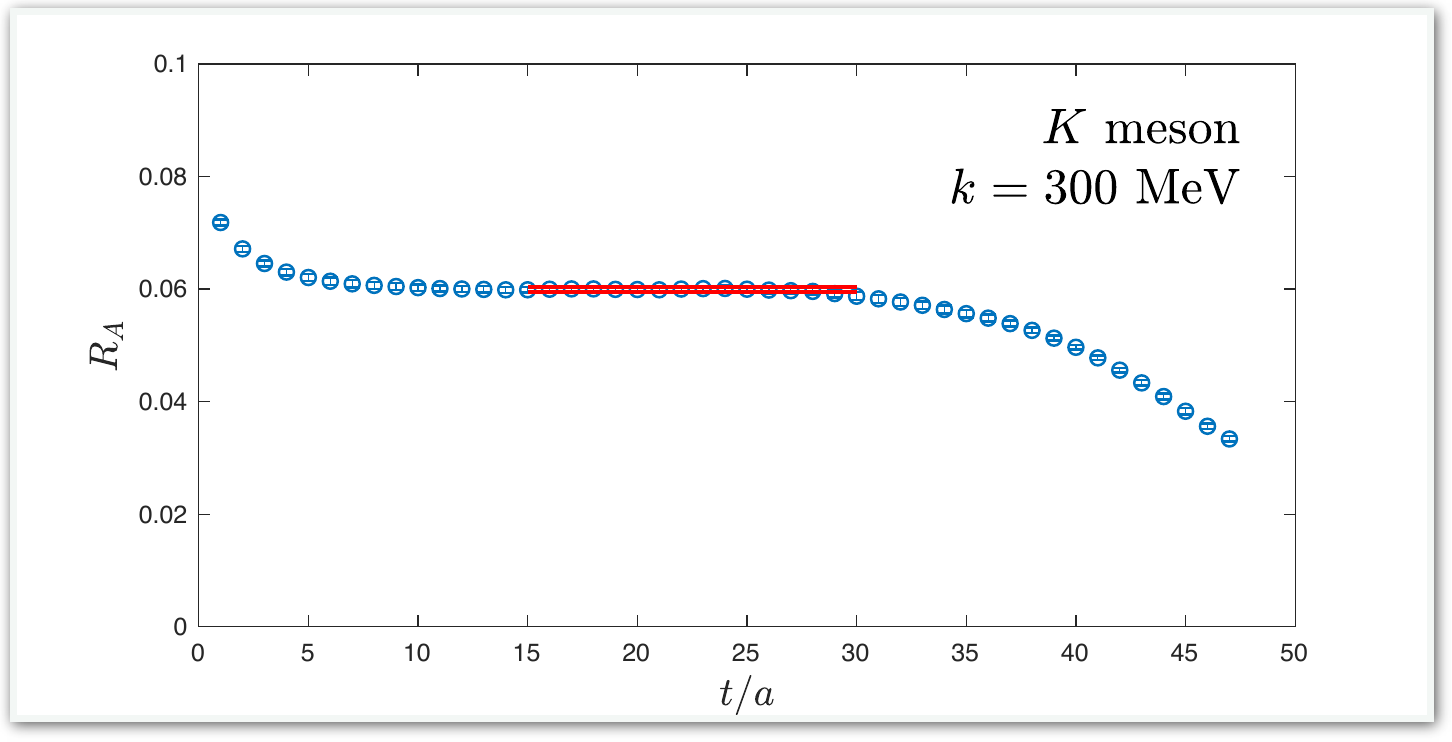}\hfill
\includegraphics[width=0.45\textwidth]{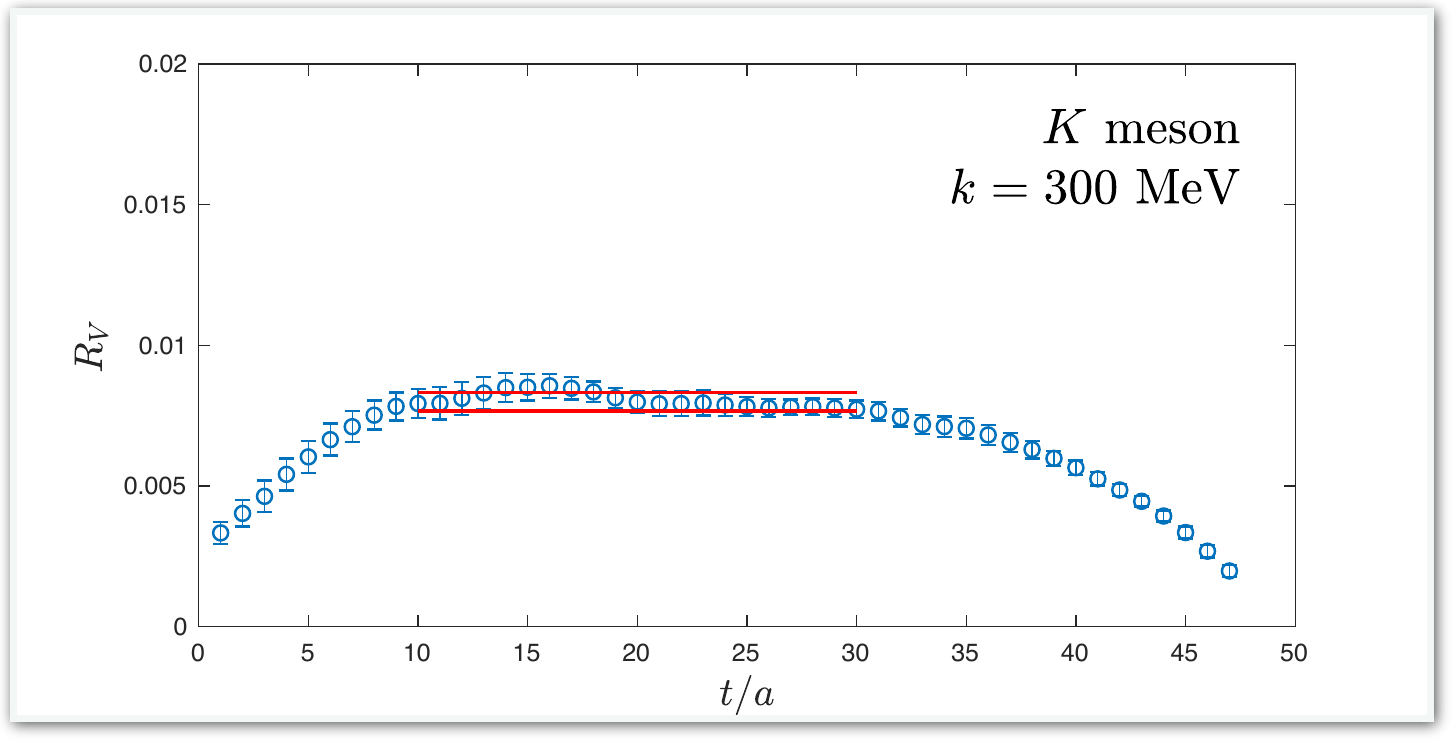}\\
\includegraphics[width=0.45\textwidth]{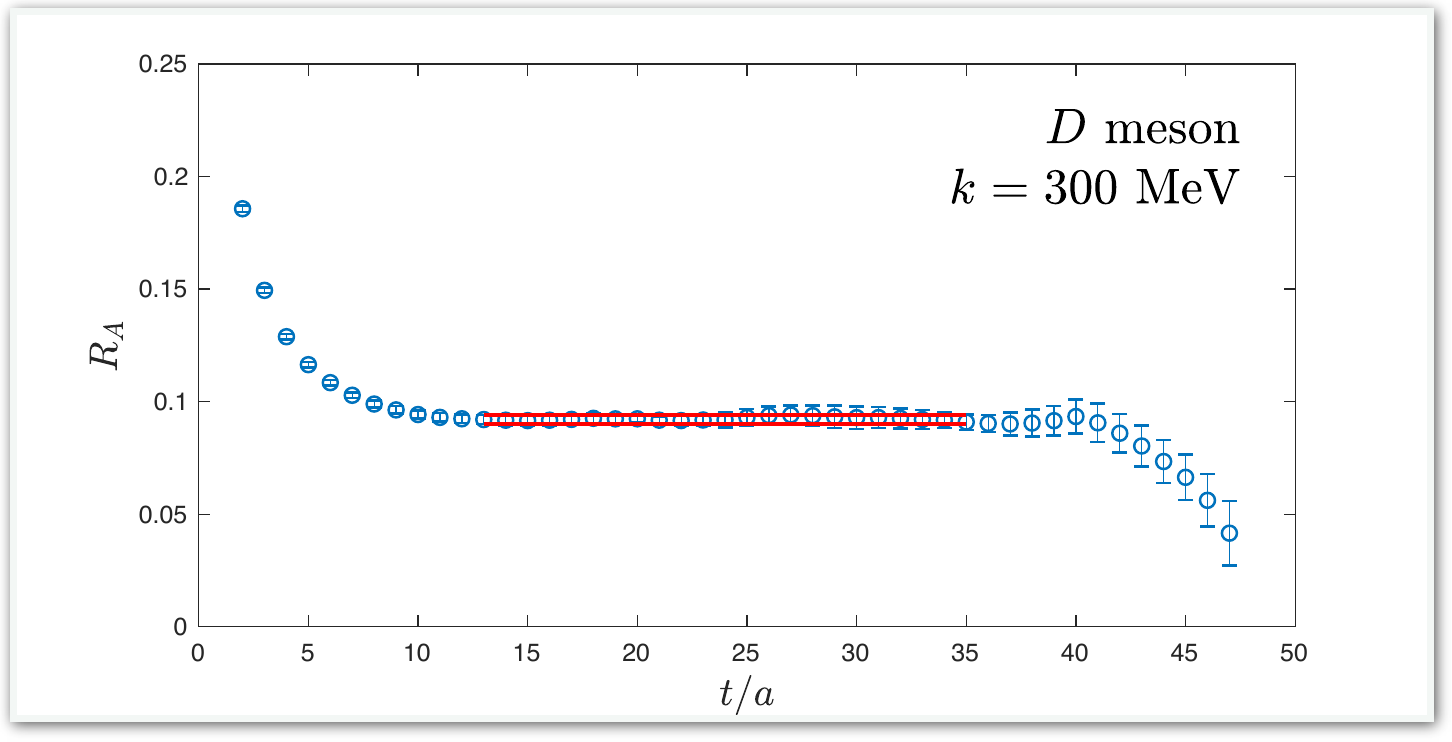}\hfill
\includegraphics[width=0.45\textwidth]{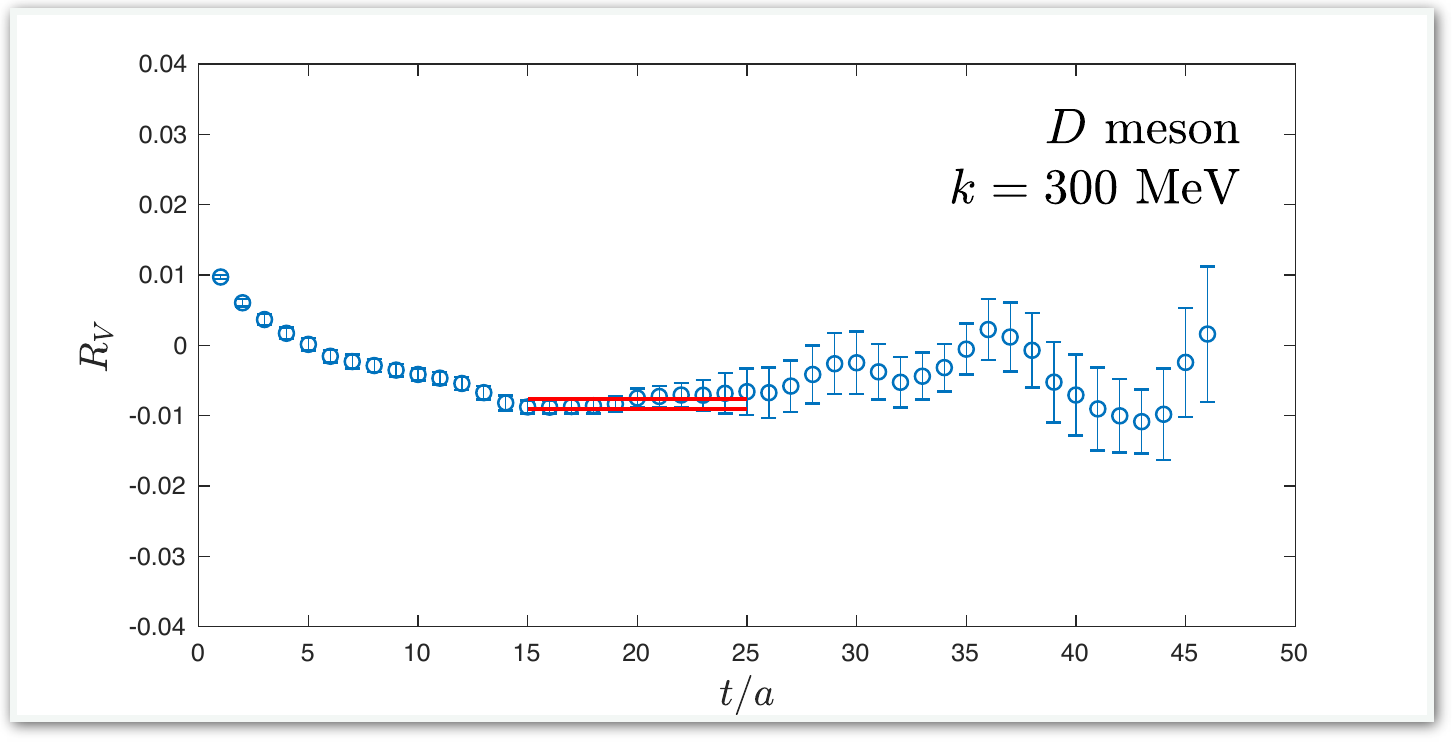}
\end{center}
\caption{
\footnotesize  Examples of plateaux fits for the ratios $R_A(t,T/2)$ (left) and $R_V(t,T/2)$ (right).}
\label{fig:plateaus}
\end{figure}
For $t \gg 0$ we get the following  numerical estimators for the form-factors 
\bea &&
R_A(t) =
\frac{m_P}{4 p\cdot k}\sum_{r=1,2}\sum_{j=1,2} \frac{R^{j r}_A(t;\vec p,\vec k)}{\epsilon^j_r}\to \,  \left[F_A(x_\gamma)+\frac{2f_P}{m_P x_\gamma}\right]  \;,
\nonumber  \\
&&
R_V(t)=
\frac{m_P}{4}\sum_{r=1,2}\sum_{j=1,2} \frac{R^{jr}_V(t;\vec p,\vec k)}{
i \left(E_\gamma\, \vec \epsilon_r \wedge \vec p-E\, \vec \epsilon_r \wedge \vec k\right)^j}\to \, F_V(x_\gamma)\;. 
\label{eq:estimators}
\eea
%
At  finite $T$, by using the  formulae above  which are valid for $t>0$, we fit the ratios $R_{A,V}(t)$ by searching a plateau in the region $0 \ll t \ll T/2$ .   We also exploit time-reversal symmetries  to include the plateaus  of  $R_{A,V}(t)$ obtained at  $t>T/2$.      The values of the meson energies and of the matrix element $\bra{P(\vec   p)}P\ket{0}$ needed to build these estimators are  obtained from standard effective-mass/residue analyses of pseudoscalar-pseudoscalar two-point functions.  The  pseudoscalar-axial two-point function  is used to  extract the decay constants $f_P$  in  order  to separate  $F_A$ from the point-like contribution $2f_P/(m_P x_\gamma)$.
\begin{figure}[!t]
\begin{center}
\includegraphics[width=0.45\textwidth]{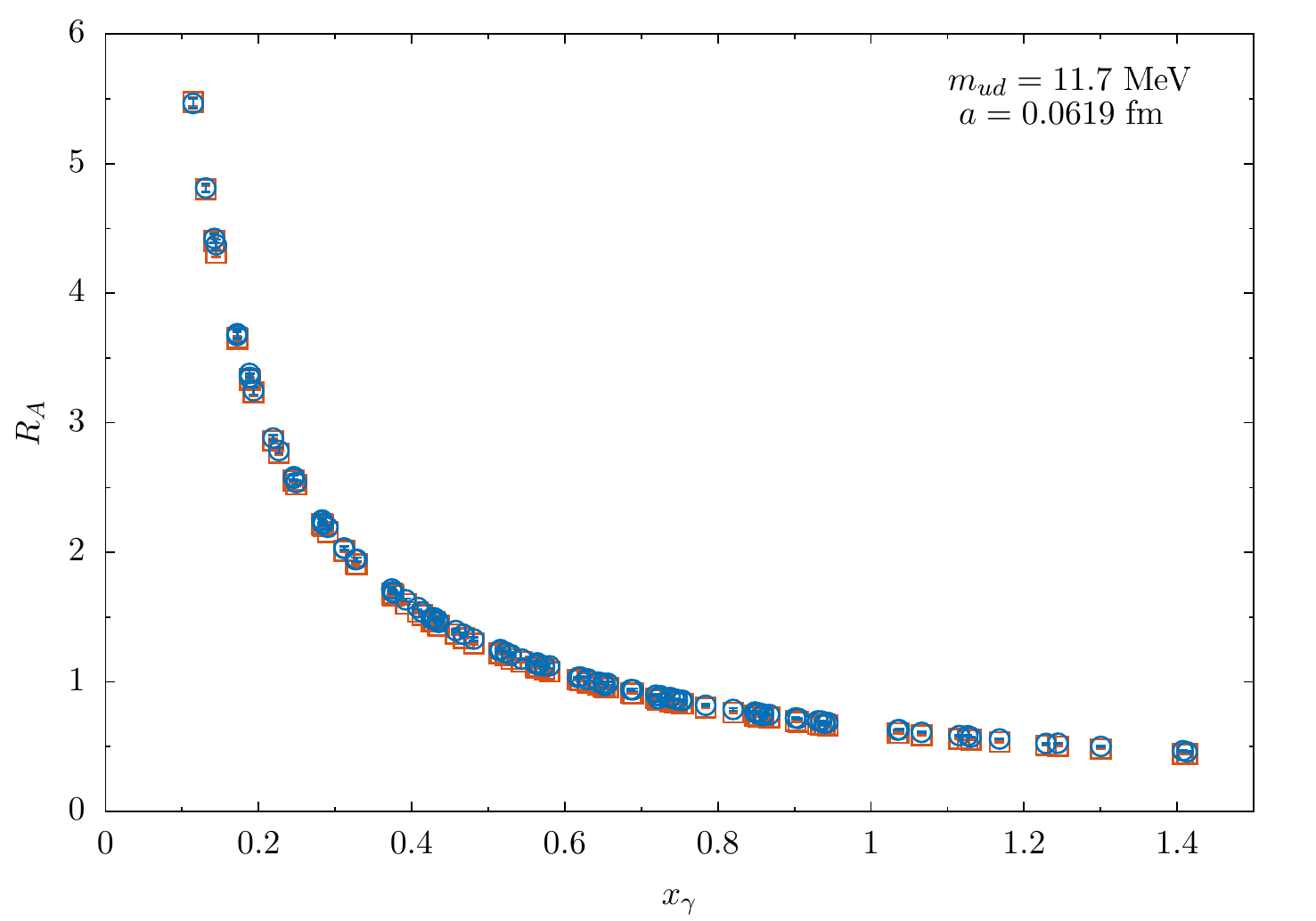}\hfill
\includegraphics[width=0.45\textwidth]{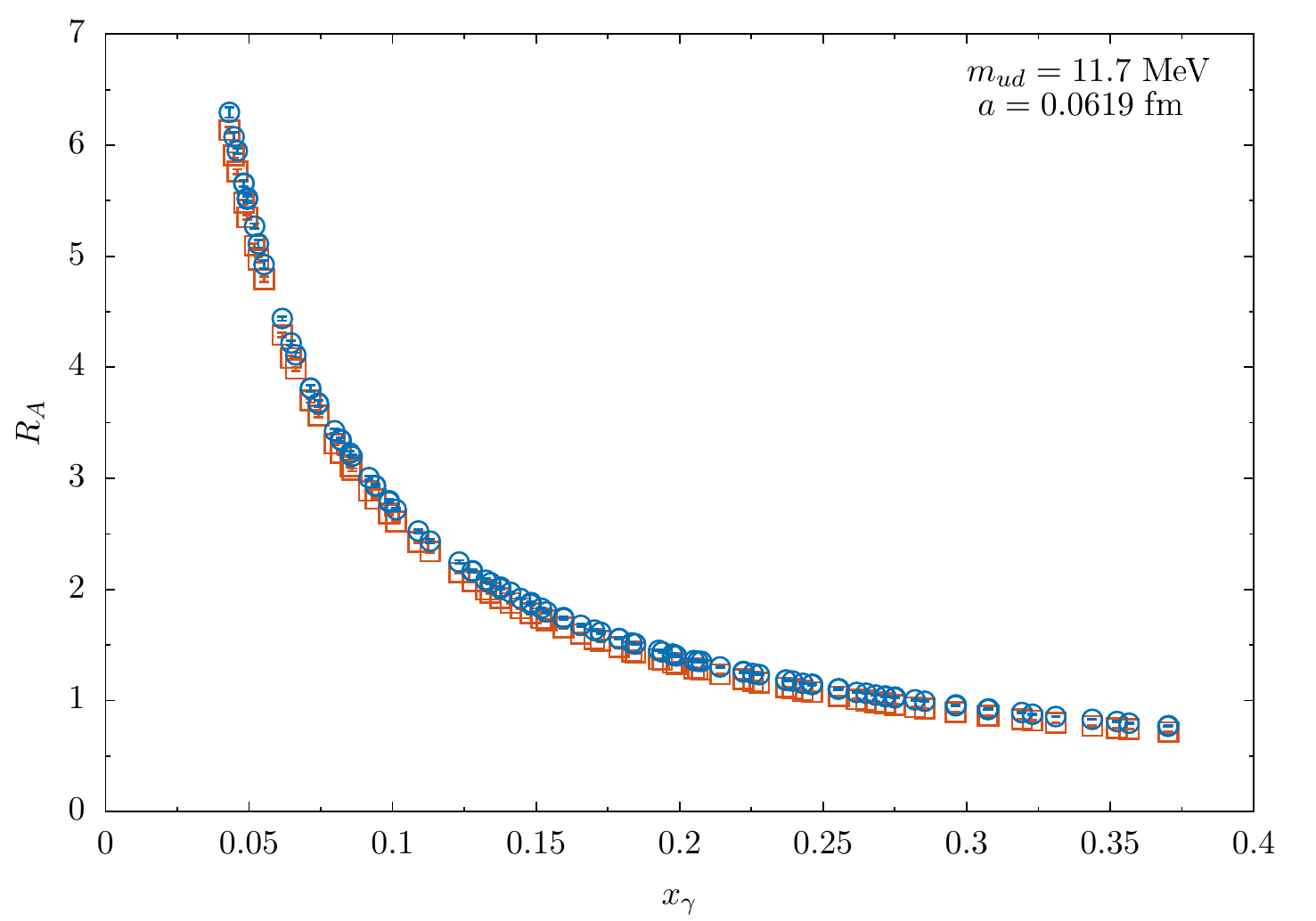}
\end{center}
\caption{\footnotesize The extracted value of $R_A(x_\gamma)$, Eq.~(3.5), as a function of $x_\gamma$ for the $K$ meson (left) and for the $D_s$ meson (right). The (red)  squares  represent the point-like  contribution given by  $2f_P/(m_P x_\gamma)$.}
\label{fig:AA}
\end{figure}

\begin{figure}[!t]
\begin{center}
\includegraphics[width=0.45\textwidth]{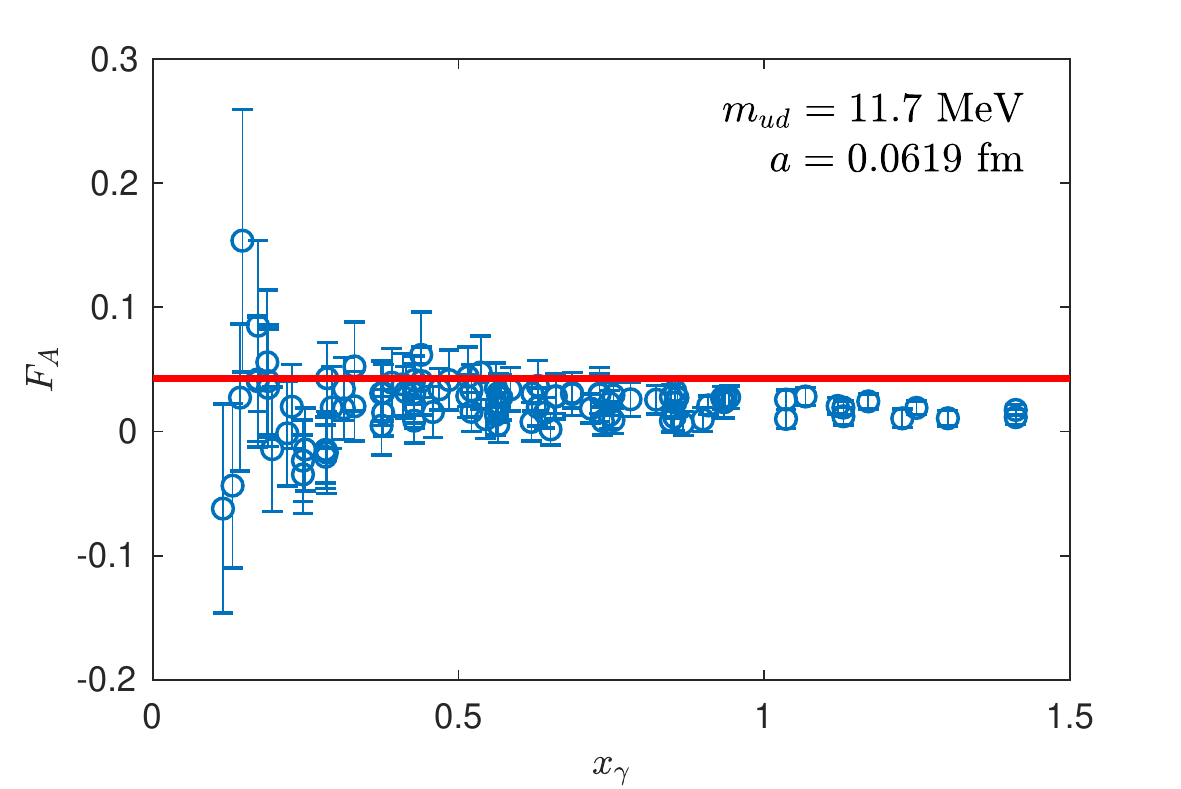}\hfill
\includegraphics[width=0.45\textwidth]{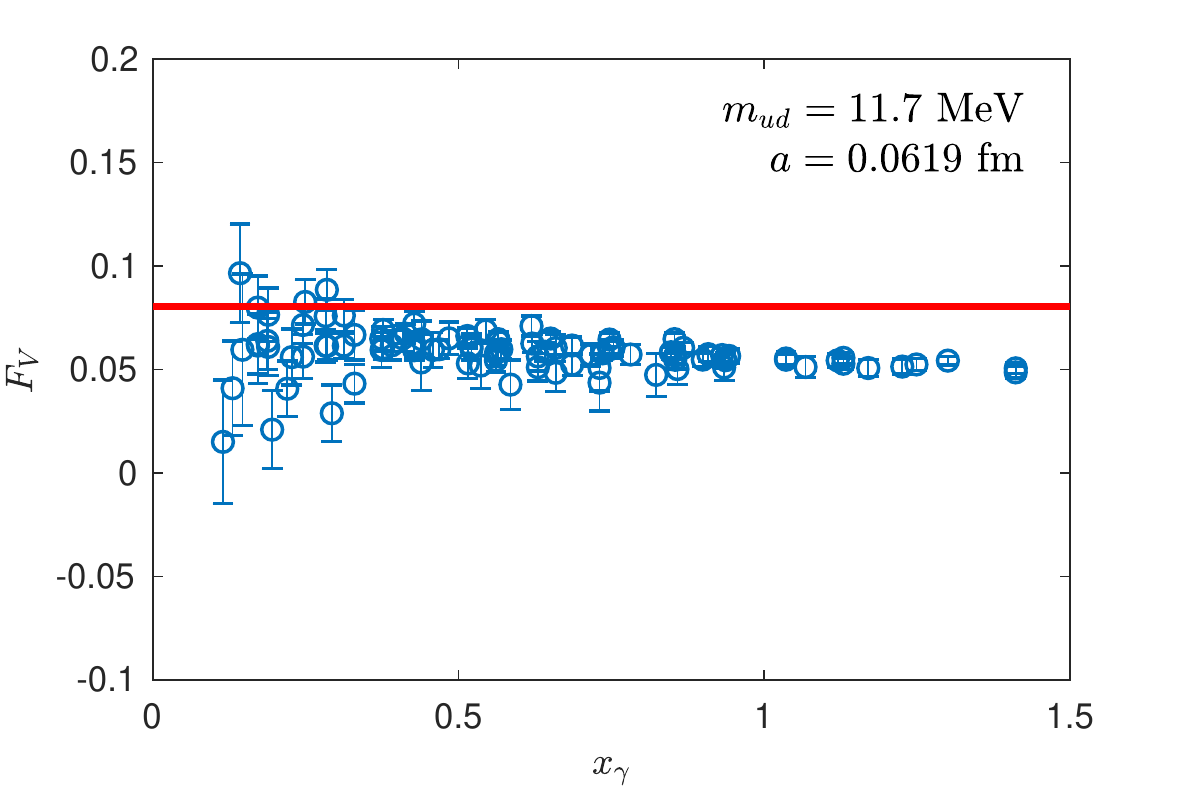}
\end{center}
\caption{\footnotesize  The extracted value of the kaon form factors $F_A(x_\gamma)$ (left) and $F_V(x_\gamma)$ (right) as a function of $x_\gamma$. The (red)  lines correspond to the $\chi PT$  predictions obtained by using the formulae discussed in the text.}
\label{fig:FA}
\end{figure}
\section{Numerical results}
\label{sec:numerical}
All the results presented in this section are preliminary. We have used the  gauge configurations given in table II of ref.~\cite{DiCarlo:2019thl},  produced with $2+1+1$ twisted mass fermions at three different values of the lattice spacing,  $a [{\rm fm}]=0.0085(36),0.00815(30),0.0619(18)$,  with meson masses in the range $250$-$1930$~MeV.  In total we have included 100 
different combinations of momenta obtained by assigning to each of the $\theta_{i=0,t,s}$ five different values; making  the same assignements for all  choices of the quark masses. 
All the plots  below correspond to the case of $K$ and $D_{(s)}$ mesons  at  unphysical values of the  $\overline{\rm MS}$ renormalised light-quark mass, $m_{ud}(2$~GeV$)=11.7$~MeV, and have been obtained from a simulation at $a=0.0619$~fm.  Thus the reference meson  masses  are $ M_D=1933\,(50)$~MeV,
$M_K=535\,(14)$~MeV and $M_\pi= 255\,(7)$~MeV.   Similar plots can be shown for other values of the simulated parameters.  In Fig.~\ref{fig:plateaus} we show examples of plateaux for the ratios $R_{A,V}(t)$ for the $K$ and $D$ mesons. This figure is representative of the signal quality, also for other values of masses and momenta. In Fig.~\ref{fig:AA} we show the extracted value of $R_A(x_\gamma)$, Eq.~(\ref{eq:estimators}), for the $K$ meson and for the $D_s$ meson. In both cases  the point-like contribution, corresponding to the term $2f_P/(m_P x_\gamma)$ dominates the form factor.  From the measured decay constant and mass, we can subtract the point-like term and extract $F_A(x_\gamma)$.  In the left-hand plot of  Fig.~\ref{fig:FA} we show  $F_A(x_\gamma)$ as a function of $x_\gamma$ and compare it to the lowest non-trivial order in chiral  perturbation theory $\chi PT$, given by $F_A(x_\gamma)= {\rm const.} = 8 m_K (L_9^r+L_{10}^r)/f_K$, indicated by 
a  line with $L_9^r+L_{10}^r\simeq 0.0017$~\cite{Bijnens:2014lea}.  On the right hand plot  of Fig.~\ref{fig:FA} we can compare the directly computed value of  $F_V$ to its $\chi PT$ prediction, $F_V(x_\gamma)= {\rm const.} =  m_K /(4\pi^2 f_K)$.  In a first exploratory study we covered the full  physical range of  $x_\gamma$  in the kaon case (indeed we even have data for  unphysical values corresponding to $x_\gamma > 1$) and for  $0\le x_\gamma\le 0.4$, corresponding to $E_\gamma \lesssim 400$~MeV,  for the $D_s$ meson.  We are currently improving our lattice data and, after a detailed analysis of all the systematics,  we shall provide first-principles phenomenologically relevant results for the form factors in the full kinematical range  for both  light and  heavy  mesons. 
The form factors for heavy mesons will represent in this respect a totally  unexplored field of investigation while,  in the case of light mesons, our first-principle results will make it possible to avoid  $\chi PT$ in phenomenological analyses.

In conclusion we have shown that, with  moderate statistics, it is possible to extract with good precision  the form factors relevant for $P \to \ell \bar \nu_\ell \gamma$ decays for both  light and  heavy mesons   and  that it is possible to study their momentum dependence.  In the near future we will be able to compare the precise theoretical predictions with  experimental measurements.
\begin{acknowledgments}
We acknowledge use of CPU time on Marconi-KNL at CINECA within the PRACE project Pra17-4394. 
 V.L., G.M. and S.S. thank MIUR  (Italy)  for  partial  support  under  the  contract  PRIN  2015. 
C.T.S.  was  supported by an Emeritus Fellowship from the Leverhulme Trust.  
N.T.  acknowledges the University of Rome Tor Vergata for the support granted to the project PLNUGAMMA.
\end{acknowledgments}

\end{document}